\documentclass[sn-mathphys,Numbered]{sn-jnl}
\usepackage{graphicx}%
\usepackage{multirow}%
\usepackage{amsmath,amssymb,amsfonts}%
\usepackage{amsthm}%
\usepackage{mathrsfs}%
\usepackage[title]{appendix}%
\usepackage{xcolor}%
\usepackage{textcomp}%
\usepackage{manyfoot}%
\usepackage{booktabs}%
\usepackage{algorithm}%
\usepackage{algorithmicx}%
\usepackage{algpseudocode}%
\usepackage{listings}%
\usepackage{lineno}
\usepackage[caption=false]{subfig}

\begin{document}
\title[Article Title]{\parbox{\dimexpr\textwidth+0cm\relax}{\centering Chemist-X: Large Language Model-Powered Agent for Recommending Reaction Conditions in Chemical Synthesis and Autonomous Laboratories}}

\author[1]{Kexin Chen}\equalcont{These authors contributed equally to this work.}
\author[2,3]{Jiamin Lu}\equalcont{These authors contributed equally to this work.}
\author[4]{Junyou Li}
\author[5]{Xiaoran Yang}
\author[5]{Yuyang Du}
\author[4]{Kunyi Wang}
\author[2]{Qiannuan Shi}
\author[4]{Jiahui Yu}
\author[4]{Lanqing Li}
\author[4]{Jiezhong Qiu}
\author[2,3]{Jianzhang Pan}
\author[2,3]{Yi Huang}
\author*[2,3]{Qun Fang}\email{fangqun@zju.edu.cn}
\author[1]{Pheng Ann Heng}
\author*[4]{Guangyong Chen}\email{gychen@zhejianglab.com}

\affil[1]{Department of Computer Science and Engineering, The Chinese University of Hong Kong, Hong Kong SAR, China}
\affil[2]{Institute of Intelligent Chemical Manufacturing and iChemFoundry Platform, ZJU-Hangzhou Global Scientific and Technological Innovation Center, Hangzhou, China}
\affil[3]{Institute of Microanalytical Systems, Department of Chemistry, Zhejiang University, Hangzhou, China}
\affil[4]{Zhejiang Lab, Hangzhou, China}
\affil[5]{Department of Information Engineering, The Chinese University of Hong Kong, Hong Kong SAR, China}

\abstract{Recent AI research plots a promising future of automatic chemical reactions within the chemistry society. This study proposes Chemist-X, a comprehensive AI agent that automates the reaction condition optimization (RCO) task in chemical synthesis with retrieval-augmented generation (RAG) technology and AI-controlled wet-lab experiment executions. To begin with, as an emulation on how chemical experts solve the RCO task, Chemist-X utilizes a novel RAG scheme to interrogate available molecular and literature databases to narrow the searching space for later processing. The agent then leverages a computer-aided design (CAD) tool we have developed through a large language model (LLM) supervised programming interface. With updated chemical knowledge obtained via RAG, as well as the ability in using CAD tools, our agent significantly outperforms conventional RCO AIs confined to the fixed knowledge within its training data. Finally, Chemist-X interacts with the physical world through an automated robotic system, which can validate the suggested chemical reaction condition without human interventions. The control of the robotic system was achieved with a novel algorithm we have developed for the equipment, which relies on LLMs for reliable script generation. Results of our automatic wet-lab experiments, achieved by fully LLM-supervised end-to-end operation with no human in the lope, prove Chemist-X’s ability in self-driving laboratories.}
\maketitle

\section{Introduction}\label{sec1}
Recent ``artificial intelligence (AI) for chemistry" research aims to free human labor with AI-supervised chemical reaction platforms \cite{Ref001, Ref002, Ref023, Ref003}. These innovative platforms employ machine learning (ML) algorithms and data-driven methods to orchestrate synthetic routes and identify optimal reaction conditions for maximizing yields. In the foreseeable future, chemical robots equipped with integrated AIs will autonomously design and refine chemical reactions, as well as automatically perform chemical reactions, allowing human experts to be freed from repetitive and time-consuming experimental operations and focus on more creative and foundational research. Yet, traditional AI systems fall short of human experts who possess a broad spectrum of chemical knowledge and can continually update their understanding by consulting related literature when dealing with new problems. Conventional AI models, being confined to the knowledge from their training data, exhibit limited generalization ability when encountering unknown reactions. These AIs can provide chemists with preliminary assistance for familiar reactions but cannot realize the ambitious aim of fully automated synthesis of unprecedented products.

Recent technical breakthroughs in retrieval-augmented generative AI (RAG-AI) have shed light on this problem. RAG-AI, as its name suggests, harnesses the strengths of retrieval-based methods and the capacities of generative AIs like large language models (LLMs). RAG enables an AI to retrieve relevant information from a vast and continuously updated knowledge source \cite{Ref004, Ref005}, while LLM equips the AI with close-to-human abilities in information analysis and code writing \cite{Ref006, Ref014}. This paper develops a sophisticated chemical reaction agent, namely Chemist-X, with RAG-AI technology. We focus on RAG for reaction condition optimization (RCO), which is equally important as retrosynthesis but receives less attention in synthetic research.

\begin{figure}[htbp]%
\centering
\includegraphics[width=0.85\textwidth]{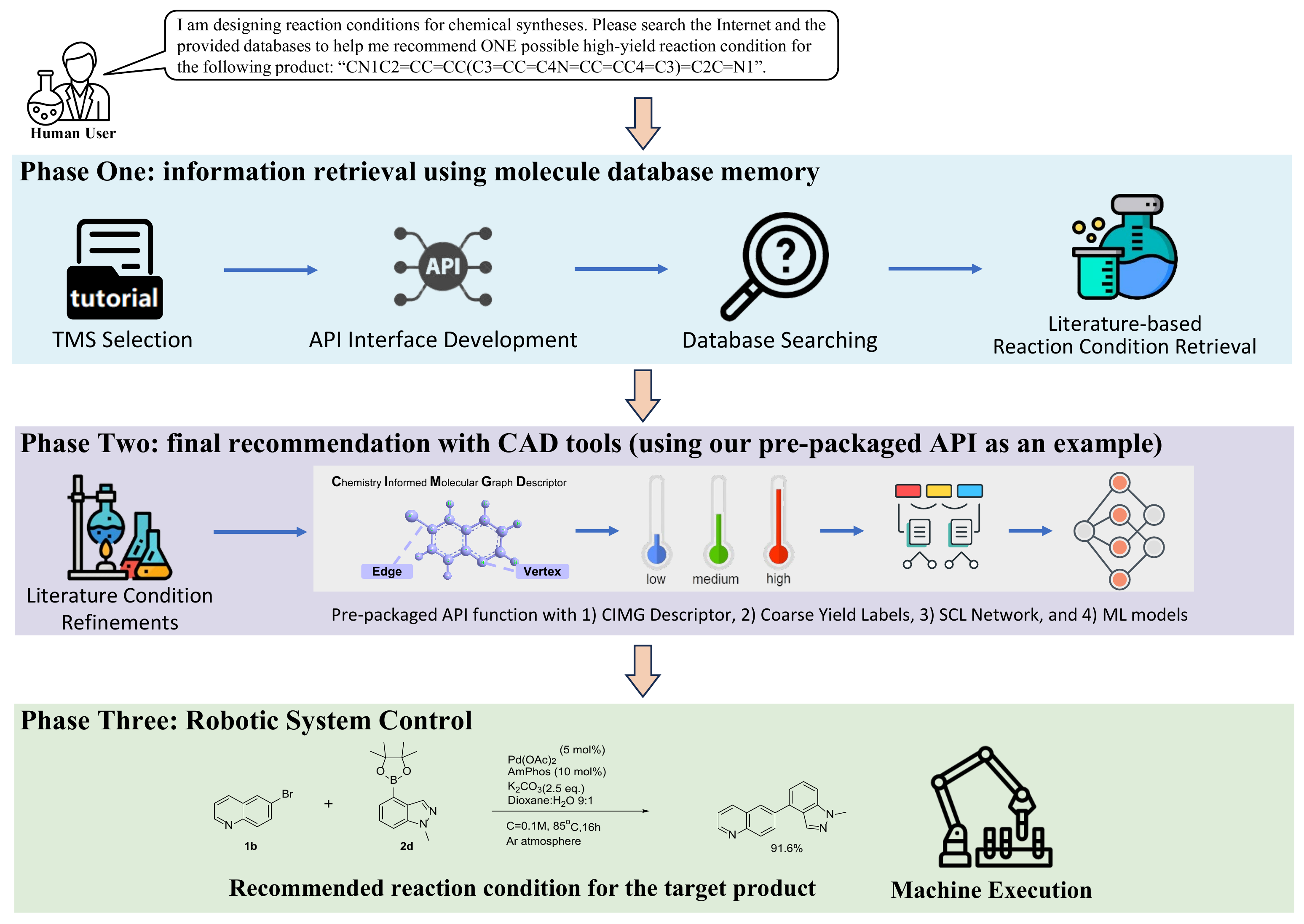}
\caption{The three-phase RCO framework of Chemist-X, in which all phases are automatically executed under the control of the LLM agent.}\label{fig1}
\end{figure}

We have designed the following three-phase pipeline to leverage LLMs’ generalization capabilities and near-human intelligence. Fig. \ref{fig1}(a) illustrates the three-phase solution with technical details, which can be summarized as follows. In Phase One, Chemist-X uses the LLM to generate usable python code for searching existing chemical literature and molecule datasets for reaction information about the target molecule (and that of its analogous molecules) to narrow down the chemical space of potential reaction conditions. Then in Phase Two, Chemist-X, as an intelligent agent supported by LLM, utilizes Computer-Aided Design (CAD) tools to select a subset of promising conditions from the refined chemical space obtained in Phase One. Finally in Phase Three, our agent controls auto-lab equipment (via LLM generated control script) to conduct wet-lab experiments to validate potential reaction conditions suggested by Phase Two. This systematic ``retrieve-recommend-validate" pattern applied in our system well aligns with the methodology a human expert might undertake when addressing the RCO challenge associated with an unfamiliar molecule, and we will later show that LLMs can well address the executive challenges in each phase owing to their strong task analysis and code generating ability. The combination of the novel three-phase pipeline and the intelligence of LLMs makes the system's reliable performance in practical tasks become possible.

We now introduce the execution of each phase. \textbf{Phase One} applies a hierarchical matching strategy to efficiently retrieve reaction conditions to a target product from both online and offline databases. Our retrieval method could be applied on various public or private databases. For concept-proving purposes, we collaborated with the Chemical Abstracts Service (CAS) organization and built a private research-use chemistry database on the Amazon Elastic Compute Cloud (EC2) platform, in which all the data has been properly licensed with CAS. Hierarchical Matching Strategy begins with an exact SMILES-matching policy to check if the target compound exists in our database. If found, the reaction conditions are retrieved using the Amazon EC2 API \cite{Ref025}. If the compound is not present, the strategy employs the PubChem API \cite{Ref007} to generate a list of similar molecules. The agent searches for the most similar molecule using exact SMILES matching and retrieves its reaction conditions. If unavailable, the process continues with the next similar molecule until successful. If no suitable match is found, the agent moves directly to Phase Two.

Complementing this strategy is an LLM-empowered code generation scheme tailored to APIs such as PubChem and Amazon EC2. These online databases typically offer APIs to facilitate information retrieval and provide examples of usage in their tutorial handbooks. However, software programming based on API documents requires a related background in computer science, and it could be a time-consuming task given that each database has its own API protocols and documentation. Phase One makes such API-based information retrieval accessible to researchers regardless of their familiarity with software programming, thereby bridging the gap between chemists and computer scientists.

In pursuit of automatic API programming without human intervention, we harness in-context learning (ICL) \cite{Ref009, Ref010}, an effective few-shot learning method that equips the LLM with necessary API knowledge as background information and augments its code generation capabilities using exemplar codes in prompts. It is important to note that providing the LLM with the entire programming handbook in ICL is counterproductive, as this could overwhelm the LLM, making it struggle to pinpoint the most relevant code segments. To address this problem, we put forth the concept of top match slice (TMS), which refers to the portion of the document bearing the utmost semantic similarity to the problem description provided to the LLM. This similarity is measured by the distance between our problem description and the tested document slice within the semantic embedding space \cite{Ref011}. Our experiments indicate that incorporating the TMS of an API handbook into the ICL prompt significantly enhances the LLM's performance in information retrieval. The scheme outperforms ICL with complete documentation or random document slices in multiple aspects, including higher code reliability, better cost-efficiency, and reduced response time.

\textbf{Phase Two} makes the recommendations. Once the chemical reaction subspace has been refined with the effort in phase one, our focus shifts to identifying the optimal set of reaction conditions with top yield expectations. There are some prior studies investigating CAD schemes for such recommendation tasks \cite{b25, b4, b6}. Given the intense research interest in RCO for chemical synthesis, there may be more advanced CAD schemes in the future. Chemist-X serves as a universal platform for cutting-edge CAD algorithms, wherein the latest research findings could be integrated in the form of APIs. To fulfill this objective, our agent harnesses LLMs to comprehend the functionality of each chemical CAD algorithm via its documentation and to select appropriate CAD tools through API programming. As the chemistry community continues to update CAD tools, our platform enables the continuous evolution of Phase Two, ensuring it remains at the forefront of technological advancement.

Apart from the general CAD platform provided in Phase Two, this paper developed an advanced ML model for the RCO task. We packaged the algorithm into an API so that the agent can execute the new model for concept proving. Specifically, in our new model, we designed a novel chemical reaction fingerprint and applied it to enhance the conventional ML models. The fingerprint diverges from traditional reaction encoding approaches in two key aspects. First, we apply chemistry-informed molecular graph (CIMG) \cite{bib15}, an innovative molecular descriptor based on graph neural networks (GNNs) \cite{Ref016}, as the underlying molecule encoding method to boost the quality of molecular descriptions. Second, we integrate supervised contrastive learning (SCL) \cite{bib19} in our fingerprint-generation network, thus enhancing the model's ability to distinguish the unique features of high-yield reactions from others. Further, to better serve the SCL process, we replaced the noise-sensitivity numerical descriptions of reaction yields with coarse categorical yield labels. With the above efforts, our reaction embedding scheme, denoted as the coarse-label SCL (CL-SCL) fingerprint, demonstrates exceptional efficacy in capturing the essential characteristics of high-yield reactions. It constantly outperforms prior studies in terms of precision and robustness across a variety of ML models.

\textbf{Phase Three} introduces a general computer control method that enables our agent to operate wet-lab experimental platforms to validate the reaction conditions recommended in Phase Two. Our motivation is that many wet-lab experimental platforms do not have predefined APIs. Instead, many rely on user interactions (e.g., mouse clicks or touchscreen inputs) within software interfaces for operation. To enhance the generalization capability of our AI agent, we designed a general computer control framework that can simulate human manipulation and learn from human-operated demonstration videos. This is achieved by utilizing the LLM's reasoning ability, its code generation ability (based on the \textit{pyautogui} Python package), and the assistance of several tools tailored to the agent: a key frame detection algorithm, a mouse-click detection algorithm, and an optical character recognition scheme.

This paper implements the automation control algorithm developed in Phase Three on a large-scale molecular manufacturing platform known as iChemFoundry (IC). With advantages such as full automation, time efficiency, reduced risk, high reproducibility, and broad versatility, the IC platform effectively addresses the widespread challenges of automating chemical synthesis in laboratory settings \cite{bib28, bib29}. Notably, the IC platform successfully automates reaction pretreatment, including solid dosing and synthesis under inert gas protection—two critical challenges faced by currently reported robotic systems \cite{bib30}.

For testing, we use the Suzuki–Miyaura coupling, a representative chemical reaction in industrial chemistry, as the target reaction. During the execution of these Suzuki–Miyaura reactions, Chemist-X recommends specific reaction conditions, which are then automatically validated on the IC platform, utilizing Unchained Labs, the HPLC characterization island, and integrated robotic arms. The IC platform autonomously executes all experimental operations, including pretreatment with solid and liquid reagents, synthesis under inert gas protection, post-treatment, sample transfer, and reaction characterization, without human intervention. The compelling results from wet-lab experiments demonstrate the feasibility of using LLMs to replicate human interactions with software interfaces for general equipment control.

In general, our three-phase framework supervised by LLM, along with the specific techniques employed in each phase, enables Chemist-X to automatically acquire the latest online molecules/literature knowledge, utilize the most cutting-edge CAD tools, and control automatic wet-lab equipment for reaction validation. With these advantages, Chemist-X makes a significant contribution to the AI-chemistry community and makes a meaningful step towards the ultimate goal of AI-supervised chemical reaction platforms.

\section{Results and Discussions}\label{sec2}
This section demonstrates how Chemist-X completes the RCO task with an example Suzuki reaction targeting \textit{6-(1-methyl-1H-indazol-4-yl) quinoline}, whose SMILES is written as CN1C2=CC=CC(C3=CC=C4N=CC=CC4=C3)=C2C=N1. We now investigate the agent's performance in Phase One and Phase Two via comprehensive unit tests, and then we validate the generated recommendation results via wet-lab experiments controlled by Phase Three.

\subsection{Phase One Unit Tests}
\begin{figure}[htbp]
\centering
\includegraphics[width=0.9\textwidth]{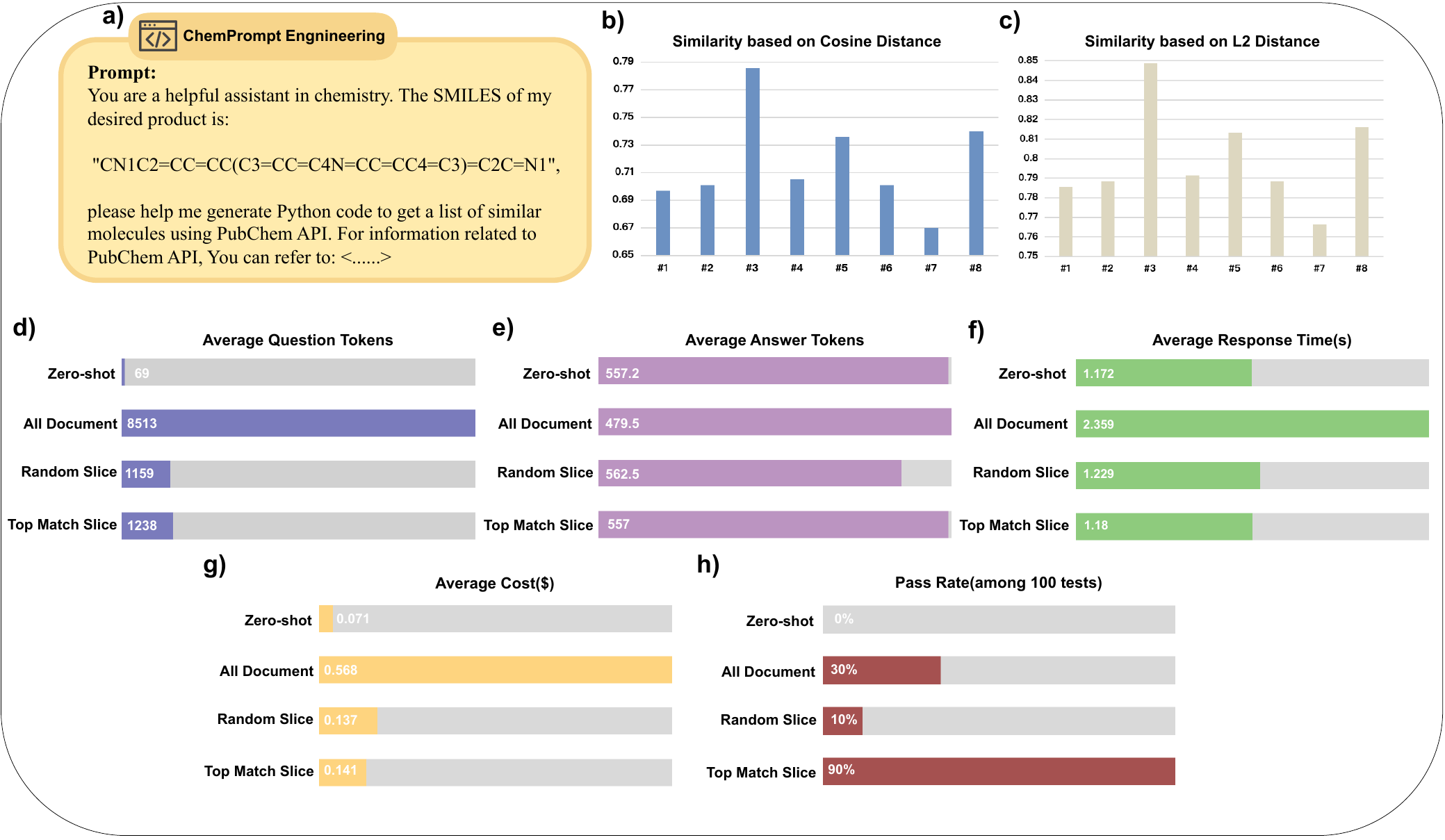}
\caption{Experimental details of Phase One's unit test. \textbf{a) }The prompt we used for the auto code generation in this experiment. Here $<...>$ is a placeholder representing the flexible part of the prompt to be selected with the TMS scheme. \textbf{b) and c) }Similarity scores between each documentation slice and the fixed part of our prompt. The scores in b) and c) are calculated with cosine distance and L2 distance, respectively. \textbf{d) to h) }Compression of four different prompting schemes (zero-shot/all-document/random-slice/TMS prompting) in terms of cost and performance.}\label{Fig_PhaseOne}
\end{figure}

Phase One employs ICL to craft Python code capable of interfacing with Amazon EC2 and PubChem via its official APIs. Since the code generation pipeline for the Amazon EC2 platform is highly similar to those of the PubChem API, this subsection uses the PubChem API for demonstration to fulfill the page limit. Figure \ref{Fig_PhaseOne}a illustrates the prompt we used for the code generation pipeline, which includes a fixed part and a flexible part. Our question is described in the fixed part, while the reference example of ICL should be contained in the flexible part. Our agent identifies the most pertinent section of the API documentation and uses it to fill in the flexible part of the prompt.

During our experiment, Chemist-X splits the text of PubChem's API documentation \cite{Ref007} into eight discrete slices with similar lengths. It then employs \textit{text-embedding-ada-002} (henceforth referred to as ADA-002), an advanced natural language processing (NLP) model from OpenAI \cite{Ref018}, to convert these document slices and our query prompt into vector representations within a semantic space. After semantic embedding, we measure the semantic similarity between the vector of each document slice and that of our prompt. For a comprehensive evaluation, we test two classical semantic similarity metrics in NLP research: Cosine Similarity \cite{Ref019} and L2 Similarity \cite{Ref020}. For a detailed explanation of the embedding procedure and the two metrics tested, we refer readers to the methodology section.

If a slice exhibits high semantic similarity to our questioning prompt, it is likely relevant to our task and could be considered a potential candidate for completing the missing part of the prompt shown in Figure \ref{Fig_PhaseOne}a. Our experimental findings, depicted in \ref{Fig_PhaseOne}b and \ref{Fig_PhaseOne}c, confirm this argument: across all eight candidates, slice \#3 achieves the highest semantic similarity regardless of the metric applied. Upon manually reviewing PubChem's documentation, this slice indeed pertains to the ``similar molecules inquiry function" provided by PubChem and includes a practical example that guides users on how to implement the inquiry function through programming. Therefore, the agent treats slice \#3 as the TMS and uses it to fill in the prompt, eventually making the LLM perform more effective ICL.

The agent's accuracy and efficiency are substantially improved by incorporating the TMS. We compare our TMS-ICL approach against three alternatives: (i) zero-shot learning without any background information or examples, (ii) ICL using the complete API documentation as background reference, and (iii) ICL with a randomly chosen slice as background reference. We conduct ten trials of our TMS-ICL and the three alternative methods using a GPT-4 model. We assess the performance of these four code-generation strategies based on computational resources used and the success rate in generating functionally correct code in the ten trials. Specifically, we consider four aspects of computational requirements: the average number of question/answer tokens used in one interaction with the GPT-4 model, the model's average response time in completing one code design task, and the average cost of utilizing the commercial LLM model for one time. Experimental results presented from Figure \ref{Fig_PhaseOne}d to Figure \ref{Fig_PhaseOne}h indicate that our TMS-ICL method consumes slightly more computational resources than the random slice approach yet delivers a remarkably higher success rate than all three alternatives. The zero-shot approach uses the least computational resources but has a disappointingly low success rate. The all-document approach shows modest improvement over zero-shot learning but does not match the performance of our TMS-ICL and requires substantially more computational resources. 

\subsection{Phase Two Unit Tests}
Phase Two develops a novel RCO algorithm with CL-SCL fingerprint, as a proof of concept, to show Chemist-X's ability to utilize updated CAD tools. Unit tests in this subsection focus on the efficacy of the proposed CAD scheme, while a systematic validation of the agent is presented in the next subsection via wet-lab experiments. To examine the generalization ability and the universality of the fingerprint, experiments are conducted as follows: we use a Buchwald-Hartwig reaction dataset \cite{b4} for model training and test the model with a Suzuki-Miyaura reaction dataset \cite{Ref021}. Further, in the evaluation, we follow the traditions established in prior RCO works \cite{b6} and use the maximum observed yield as the figure of merit. We denote the maximum observed yield of final recommendations by $\mu_N$, where $N$ is the batch size BS of the final recommendation.

\textbf{Benchmarking Experiments:} We first compare our CL-SCL fingerprint against two reaction fingerprints from prior studies: DRFP \cite{bib24} and Mordred \cite{bib25}. We validate these fingerprints with three ML algorithms, i.e., random forest (RF) \cite{bib22}, tabular transformer (TT) \cite{Ref024}, and Xgboost (XGB) \cite{bib23}, under different batch size configurations. Table \ref{Table_PhaseThree_Table1} presents the resulting $\mu_N$ across different experimental setups, which reveals that our CL-SCL fingerprint consistently delivers superior reaction condition recommendations than other alternatives regardless of batch size setting and the ML model harnessed.
\begin{table}[!ht]
\centering
   \subfloat[Benchmarking Results ($\mu_N$ is recorded in the form of percentage, and BS refers to batch size) \label{Table_PhaseThree_Table1}]{
     \begin{tabular*}{1\textwidth}{@{\extracolsep{\fill}}ccccccccccc@{\extracolsep{\fill}}}
\toprule
& \multicolumn{3}{@{}c@{}}{DRFP Fingerprint} & \multicolumn{3}{@{}c@{}}{Mordred Fingerprint} & \multicolumn{3}{@{}c@{}}{Our CL-SCL Fingerprint} 
\\\cmidrule{2-4}\cmidrule{5-7}\cmidrule{8-10}%
BS & TT & RF & XGB & TT & RF & XGB & TT & RF & XGB \\
\midrule
1  & 39.242 & 53.950 & 52.750 & 48.876 & 34.016 & 48.118 & \underline{\textbf{61.253}}  & 56.510 & 53.553\\
2  & 40.290 & 63.676  & 68.907  & 53.716 & 40.711 & 63.778 & \underline{\textbf{75.673}} & 68.087 & 68.937\\
3  & 70.258 & 72.153  & 72.970  & 65.806 & 77.386 & 77.796 & \underline{\textbf{86.953}} & 79.663 & 70.653\\
4  & 86.365 & 80.251 & 76.790 & 75.998 & 81.474 & 83.390 & \underline{\textbf{90.437}} & 83.109 & 74.223\\
5  & 89.346 & 88.280  & 77.400  & 85.534 & 83.588 & 86.327 & \underline{\textbf{91.363}} & 86.551 & 79.665\\
\bottomrule
\end{tabular*}\label{table1}
}\hfil
\subfloat[Experimental Results of Ablation Study ($\mu_5$ is recorded in the form of percentage)\label{Table_PhaseThree_Table2}]{
\begin{tabular*}{1\textwidth}{@{\extracolsep{\fill}}cccc@{\extracolsep{\fill}}}
\toprule
 & i. Remove the CL-SCL network & ii. Remove CIMG encoding & iii. Our original method \\
\midrule
TT  & 80.232 & 82.273 & \underline{\textbf{91.363}}\\
RF  & 81.186 & 76.140 & \underline{\textbf{86.551}}\\
XGB & 73.899 & 76.955 & \underline{\textbf{79.665}}\\
\bottomrule
\end{tabular*}}\hfil
\caption{A detailed result record for experiments conducted in Phase Two.}\label{Table_PhaseThree_Table}
\end{table}

\textbf{Ablation Study}: As presented in the methodology section in detail, there are two critical operations in generating the fingerprint: 1) the CIMG molecule encoder and 2) the CL-SCL network. A question we are interested in is the effectiveness and necessity of each operation. To address this query, we conducted an ablation study that tests the indispensability of each element. The ablation study sets the batch size as 5. In Experiment (i), we remove the CL-SCL network, allowing the chemical reactions encoded by CIMG to directly interface with subsequent ML models. In Experiment (ii), we substitute the CIMG encoder with a trivial encoder that simply concatenates molecule encodings. Beyond these alterations, the rest of the experimental setting mirrors the benchmarking experiment.

Table \ref{Table_PhaseThree_Table2} presents the resulting $\mu_5$ of each experiment. As the table suggests, our original method has a significantly higher $\mu_5$ than case (i) and case (ii), regardless of the ML model applied. This finding indicates that both the CIMG encoder and the CL-SCL network are important components in Phase Two.

\subsection{Phase Three Wet-Lab Validations}
This subsection presents experimental details about the agent-controlled web-lab experiments. We will start with our experimental setups on the IC platform, and then we will present experimental results and discuss them accordingly. For technical details on how the control is realized, we refer readers to Section 3, the methodology section, for details.

\subsubsection{Experimental Setup}
\begin{figure}[htbp]
\centering
\includegraphics[width=0.78\textwidth]{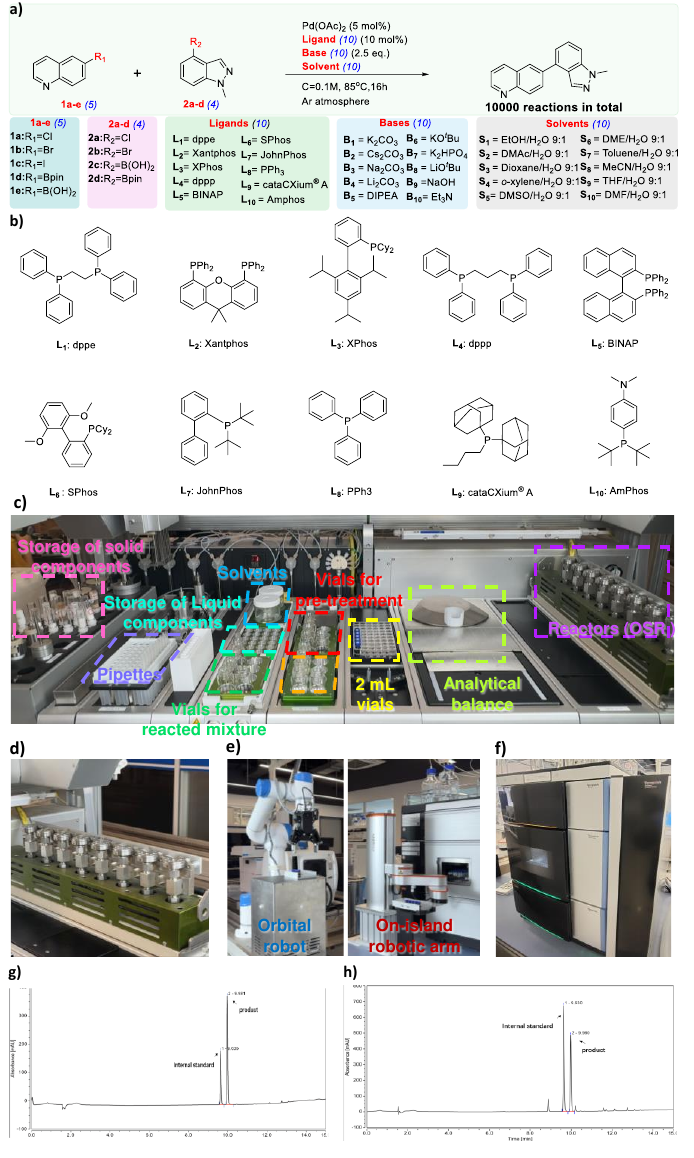}
\caption{
Overview and procedure of the studied Suzuki-Miyaura reactions. \textbf{a) }The chemical space of the studied Suzuki-Miyaura reaction. \textbf{b) }Structure of phosphorus ligands shown in a). \textbf{c) }Unchained Lab in the IC platform, equipped with modules for storage of solid, liquid reagents and solvents, pipettes for viscous or trace liquid reagents, vials for different experiments, an analytical balance and reactors (OSR). \textbf{d) }Reactors (OSR) of the Unchained Lab with inert gas protection. \textbf{e) }The orbital robot and on-island robotic arm of the IC platform for achieving automated transfer operations. \textbf{f) } Identification of the product and yield analysis with the HPLC-MS and HPLC systems. \textbf{g) }The calculation of the Relative Response Factor with the internal standard method. \textbf{h) }Yield analysis of each reaction with HPLC.}\label{fig4}
\end{figure}

To obtain 6-(1-methyl-1H-indazol-4-yl), the target product, we consider the following chemical subspace encompassing four key dimensions: reactant, ligand, base, and solvent. Further, considering the typical reaction conditions of Suzuki reactions and the availability of chemicals, the initial chemical subspace, as shown in Figure \ref{fig4}a and \ref{fig4}b, has 10,000 possible reactions within it.\footnote{The chemical subspace consists of nine reactants (\textbf{1a-1c} react with \textbf{2c-2d} and \textbf{1d-1e} react with \textbf{2a-2b}, leading to 10 combinations), ten ligands, ten bases, and ten solvents, which leads to in 10000 possible reactions.} 

After operation and analysis conducted in Phase One, our agent identifies from existing literature that K$_2$CO$_3$ could be the optimal base, and dioxane/H$_2$O could be the most suitable solvent. This insight effectively reduces the unknown variables within our chemical subspace to two dimensions: reactants and ligands. Consequently, the reaction subspace is simplified to 100 viable reaction configurations. Then, Phase Two, using the TT network as the example ML model embedded after the CL-SCL network, makes final recommendations among the narrowed reaction subspace. In a practical application, Phase Two is used only once with fixed CL-SCL and TT networks. However, to comprehensively test our scheme's robustness in the face of randomness during the training process, we re-examine Phase Two three times, with each time using a different random seed in the training process.

Additionally, we also perform random selections from the original reaction space with 10,000 reactions, which serves as a comparison benchmark to our RAG-AI agent.

\subsubsection{Experimental Details and Discussions}
We now give a detailed description of our operation in the experiments. Unless otherwise specified, all reagents and solvents were commercially available and used without further purification. All reactions were performed in the optimization sampling reactor (OSR) of the Unchained Lab, which enables synthesis reactions requiring conditions of heating, pressurization, continuous stirring and inert gas protection (\ref{fig4}c).  The product was purified by column chromatography on silica gel 200–300 mesh and identified by high-performance liquid chromatography/mass spectrometry (HPLC-MS). Fourth, the reaction yield used 4-chloro-6,7-dimethoxyquinoline as an internal standard and was analyzed by HPLC, which was determined by Thermo Scientific Vanquish Core HPLC with a binary pump, a UV detector, an Acclaim 120-C18 (3*150 mm, 3 $\mu$m), using mobile phase A: H$_2$O /B: CH$_3$CN, the flow rate is 0.5 mL/min.

A typical procedure for the synthesis process and yield analysis is as follows: Before synthesis, the solid, liquid reagents and solvents were pre-prepared in storage vials, and the air in OSR was replaced with inert gas in advance. The experiment operations of the robotic system for Suzuki-Miyaura coupling, including pre-treatment, synthesis, post-treatment and characterization of the reaction mixture, were automatically performed using the Unchained Lab and robotic arms of the IC platform without human intervention (Supplementary movie 1). First, the reaction mixture was prepared by automatically adding substrates, catalyst, ligand, base and solvent into a 20 mL vial for pre-treatment. For example, for performing the first condition experiment of the first batch in Table \ref{table3}, 6-bromoquinoline (1b, 312 mg, 1.5 mmol), 1-methyl-4-(4,4,5,5-tetramethyl-1,3,2-dioxaborolan-2-yl)-1H-indazole (2d, 387 mg, 1.5 mmol), Pd(OAc)$_2$ (16.9 mg, 0.075 mmol), potassium carbonate (B$_1$, 519 mg, 3.75 mmol) and Amphos (L$_{10}$, 39.8 mg, 0.15 mmol) in dioxane/H$_2$O (9: 1, 15 mL) were added into a 20 mL vial. Five reactions with different conditions were performed in parallel for each batch of experiments (Table \ref{table3}). Then, the reaction mixture solutions were transferred to the OSR (Figure \ref{fig4}d),  heated at 85$^o$C with continuous stirring for 16 h under nitrogen atmosphere, then cooled down to room temperature. For post-treatment, 5 mg of the internal standard (IS) was dissolved in 10 mL of methanol (MeOH), and then 100 $\mu$L of the reacted mixture was aspirated and added to the IS stock solution, shaken until thoroughly mixed and transferred to 2 mL vials to prepare the mixture solution for characterization. Then, through the cooperation of an orbital robot and an on-island robotic arm (Figure \ref{fig4}e), the vial plate, containing a batch of 5 mixture solutions obtained under different conditions, was transferred to the HPLC island for characterization (Figure \ref{fig4}f).

With the IS and the reference standard (i.e., product purified by column chromatography), the relative response factor (RRF) was determined with the following equation:
\begin{equation}
\text{RRF}=\frac{{{A}_{\text{IS}}}\times {{C}_{\text{reference}}}}{{{A}_{\text{reference}}}\times {{C}_{\text{IS}}}}
\label{equation1}
\end{equation}

where $A_{\text{IS}}$ and $A_{\text{reference}}$ are the peak areas of the IS and the reference standard obtained by the HPLC system (Figure \ref{fig4}g), $C_{\text{reference}}$ and $C_{\text{IS}}$ are the known concentrations of the reference standard and the IS.

For the quantitative characterization of the target product in the reaction mixture solution, $C_{\text{product}}$ was calculated with the equation:
\begin{equation}
{{C}_{\text{product}}}=\frac{{{A}_{\text{product}}}\times {{C}_{\text{IS}}}\times \text{RRF}}{{{A}_{\text{IS}}}}
\end{equation}

where $C_{\text{product}}$ is the concentration of the product in the reacted mixture solution to be determined, $A_{\text{product}}$ and $A_{\text{IS}}$ are the peak areas of the product and the IS obtained by the HPLC system (Figure \ref{fig4}h), $C_{\text{IS}}$ is the known concentration of the IS in the reacted mixture solution, RRF is calculated by equation \ref{equation1}.

Then the product yield of the reaction was quantified with the equation:
\begin{equation}
\text{Product yield=}\frac{{{C}_{\text{Product}}}}{{{C}_{\text{substrate}}}}\times 100\%
\end{equation}

where $C_{\text{substrate}}$ is the known concentration of the substrate added in the unreacted mixture solution. The experiments at other reactions were conducted and their mixture solutions were analyzed following procedures similar to the above steps.

Experimental results are recorded in Table \ref{table3}. All three experimental batches recommended by Chemist-X reached a $\mu_5$ exceeding 90\%. In contrast, random sampling within the four-dimension space only resulted in a $\mu_5$ around 52\%. This stark comparison demonstrates the proficiency of our agent in identifying high-yield reaction conditions amidst a broad chemical reaction space, consistently outperforming the baseline method regardless of possible randomness in the network training process.
\begin{table}[htbp]
\caption{Results of wet-lab experiments.}\label{table3}
\centering
\footnotesize
\begin{tabular}{ccccccc}
\toprule
Reaction No. & Reactant1 & Reactant2 & Ligand & Base & Solvent & Yield \\
\midrule
\multirow{5}{*}{\shortstack{Experimental\\Batch \#1}} & \textbf{1b} & \textbf{2d} & Amphos & K$_2$CO$_3$ & Dioxane: H$_2$O = 9:1 & \underline{\textbf{0.916}} \\
 & \textbf{1d} & \textbf{2b} & dppp & K$_2$CO$_3$ & Dioxane: H$_2$O = 9:1 & 0.307 \\
 & \textbf{1c} & \textbf{2d} & PPh$_3$ & K$_2$CO$_3$ & Dioxane: H$_2$O = 9:1 & 0.559 \\
 & \textbf{1c} & \textbf{2c} & Xantphos & K$_2$CO$_3$ & Dioxane: H2O = 9:1 & 0.741 \\
 & \textbf{1a} & \textbf{2d} & Amphos & K$_2$CO$_3$ & Dioxane: H$_2$O = 9:1 & 0.709 \\
\midrule
\multirow{5}{*}{\shortstack{Experimental\\Batch \#2}} & \textbf{1a} & \textbf{2d} & PPh$_3$ & K$_2$CO$_3$ & Dioxane: H$_2$O = 9:1 & 0.026 \\
 & \textbf{1e} & \textbf{2b} & dppe & K$_2$CO$_3$ & Dioxane: H$_2$O = 9:1 & 0.028 \\
 & \textbf{1c} & \textbf{2d} & cataCXium A & K$_2$CO$_3$ & Dioxane: H$_2$O = 9:1 & 0.801 \\
 & \textbf{1a} & \textbf{2c} & SPhos & K$_2$CO$_3$ & Dioxane: H$_2$O = 9:1 & 0.943 \\
 & \textbf{1a} & \textbf{2c} & Xantphos & K$_2$CO$_3$ & Dioxane: H$_2$O = 9:1 & \underline{\textbf{0.986}} \\
\midrule
\multirow{5}{*}{\shortstack{Experimental\\Batch \#3}} & \textbf{1b} & \textbf{2d} & PPh$_3$ & K$_2$CO$_3$ & Dioxane: H$_2$O = 9:1 & 0.791 \\
 & \textbf{1c} & \textbf{2c} & dppp & K$_2$CO$_3$ & Dioxane: H$_2$O = 9:1 & 0.467 \\
 & \textbf{1b} & \textbf{2c} & Xantphos & K$_2$CO$_3$ & Dioxane: H$_2$O = 9:1 & 0.836 \\
 & \textbf{1d} & \textbf{2a} & Xphos & K$_2$CO$_3$ & Dioxane: H$_2$O = 9:1 & 0.412 \\
 & \textbf{1b} & \textbf{2d} & cataCXium A & K$_2$CO$_3$ & Dioxane: H$_2$O = 9:1 & \underline{\textbf{0.965}} \\
\midrule
\multirow{5}{*}{\shortstack{Random\\Batch}} & \textbf{1c} & \textbf{2c} & Xantphos & LiO$^t$Bu & MeCN: H$_2$O = 9:1 & \underline{\textbf{0.515}} \\
 & \textbf{1b} & \textbf{2d} & SPhos & NaOH & DME: H$_2$O = 9:1 & 0.269 \\
 & \textbf{1e} & \textbf{2a} & PPh$_3$ & Cs$_2$CO$_3$ & DMSO: H$_2$O = 9:1 & 0.001 \\
 \t& \textbf{1e} & \textbf{2b} & JohnPhos & Na$_2$CO$_3$ & MeCN: H$_2$O = 9:1 & 0.351 \\
 & \textbf{1a} & \textbf{2c} & PPh$_3$ & K$_2$CO$_3$ & Toluene: H$_2$O = 9:1 & 0.388 \\
\bottomrule
\end{tabular}
\end{table}

\section{Method}\label{sec3}
\subsection{Phase One: Information Retrieval Using Molecule and Literature Databases}
Databases like PubChem and Amazon EC2 typically offer API interfaces for user queries, along with comprehensive tutorials that demonstrate the functions of these APIs through examples. In Phase One, we harness the API and its documentation with our innovative TMS-ICL strategy.

Let us start with the ICL technology. Distinct from conventional supervised learning methods, ICL does not require parameter optimization within the network. It allows an LLM to form a ``short-term memory" based on specific examples. Given prompts enriched with detailed examples, an LLM can more accurately perform tasks drawing upon the contextual information assimilated from the prompt, which facilitates its stunning performance across various NLP benchmarks.

The efficacy of ICL hinges on the selection of appropriate examples. A highly relevant example can significantly enhance the LLM performance compared to a less related one. Traditional ICL research, such as \cite{bib26b}, typically involves manual example selection tailored to a particular task. The ICL process in Phase One seeks to automate this example selection so that the system can adapt to different databases without human intervention. Otherwise, a chemist would need to manually review documentation and choose examples, which would undermine the automation objective of Chemist-X.

We put forth a quantitative method for selecting the document part most relevant to our problem defined in the prompt. Our method begins by dividing the tutorial document into slices with similar lengths. Then, we tokenize the content of a document slice and convert the contextual information within the slice into vector representations using OpenAI's ADA-002 model \cite{Ref018}. Let us denote the representing vector (also known as embedding) of slice $i$ by $A_i$. Concurrently, we process the questioning prompt with the same model and denote its embedding by $B$.

Following this transformation, we compute the cosine similarity between vector $B$ and $A_i$ with the following equation
\begin{equation} \label{eq001}
d_i = 1.0 - \frac{{\sum_{k=1}^K} A_i(k) \cdot B(k)}{\sqrt{\sum_{k=1}^K A_i^2(k)} \cdot \sqrt{\sum_{k=1}^KB^2(k)}}
\end{equation}
where $K$ is the dimension of the embedding space.

An alternative way to measure the similarity is using the L2 similarity written as
\begin{equation}\label{eq002}
d^{'}_i = \frac{1} {1 + \sqrt { {\sum_{k=1}^K} \left(A_i(k)-B(k)\right)^2} }
\end{equation}

We refer to the document slice with the highest similarity to our question as TMS. It contains the most semantically relevant contexts, and is, therefore, the most suitable slice for our ICL's example input. In Fig. 1 of the extended data, we provide an example Python code snippet that was automatically generated by GPT during Phase One for illustrative purposes.

\subsection{Phase Two: Reaction Condition Optimization with Pre-packaged Fingerprint Tool}\label{subsection3C}
It is important to note that a reaction condition emerging from Phase One may be incomplete. Although a literature platform could offer comprehensive condition information for numerous chemical reactions, it might inadvertently omit critical details in a reaction, such as reagents, catalysts, solvents, and yield. In other words, detailed reaction data from the original paper might be uncaptured in the literature platform. Chemist-X aims to accelerate chemical reaction designs and facilitate future robot-operated chemical reaction platforms that free human labor by automatically providing complete recommendations with no information loss. In light of this, at the beginning of Phase Two, we fill in the overlooked information with a comprehensive set of potential solutions in our chemical subspace.

After the information-filling process, the agent needs to identify the best possible reaction conditions with the highest expected yield. Previous works have proved that advanced ML algorithms are competent for the RCO task \cite{b4, b25, b6, bib27a}. However, one major problem we noticed in these works is that their reaction encodings, also known as reaction fingerprints, are just straightforward concatenations of encoded molecules. Although these reaction encoding schemes are easy to implement, we want to emphasize the importance of having a reaction fingerprint with additional yield information integrated into it when training an ML model for yield-predicting tasks or RCO tasks targeting optimal yields.

We now illustrate how our model training process is conducted from two aspects: the pre-processing of the reaction data, and the CL-SCL network.

\subsubsection{Reaction Data Processing}
we encode all chemicals that appeared in the reaction (e.g., reactants, ligands, base, solvent, etc.) with the CIMG descriptor introduced in \cite{bib15}. The CIMG descriptor incorporates a comprehensive set of key information about a chemical molecular, specifically nuclear magnetic resonance (NMR) chemical shifts \cite{bib16} and bond dissociation energy (BDE) \cite{bib17}. In CIMG, the NMR chemical shifts and BDEs of a molecule are features as vertex and edge of a GNN model. NMR chemical shifts can be naturally represented as properties assigned to the vertices of the graph because they pertain to specific atoms, while BDEs are associated with individual bonds and can thus be captured as edge properties. As a result, GNN can seamlessly integrate chemical information into the molecular descriptors. 

With the incorporation of these chemistry features, the CIMG descriptor becomes an informative molecular descriptor and serves as a robust foundation for the reaction embedding, in which we concatenate CIMG descriptors of each chemical component to form the numerical encoding of each reaction. In a Suzuki reaction, for example, we consider the following components: electrophile, nucleophile, catalyst, ligand, base, and solvent, and the concatenated expression of the reaction is written as
\begin{equation}
{{{x}}}_{i}={{{x}}}_{{\rm{Electrophile}}}\oplus {{{x}}}_{{\rm{Nucleophile}}}\oplus {{{x}}}_{{\rm{Pd\text{ }catalyst}}}\oplus {{{x}}}_{{\rm{Ligand}}}\oplus {{{x}}}_{{\rm{Base}}}\oplus {{{x}}}_{{\rm{Solvent}}}
\end{equation}
where $\oplus$ denotes concatenation, and elements on the right side of the equation denote the CIMG descriptors of the above-mentioned vital components of a Suzuki reaction.

\subsubsection{The CL-SCL Network}
The training data of the SCL network contains random noise, as the yield of a chemical reaction may be affected by factors including but not limited to measurement errors, material purity, and environmental factors (e.g., humidity and room temperature). These random data noises could depress the effectiveness and robustness of the yield production models. For example, if reaction A has a yield of 90\% while reaction B has a yield of 91\%, then their yield gap is even smaller than the random noise of the two chemical reactions. In such cases, it is beneficial to treat the yields of these two reactions at the same level so that the ML model can encapsulate the fundamental chemical insights and underlying reaction principles shared among similar reactions. Otherwise, focusing the ML model to learn the regression knowledge and discriminate 91\% over 90\% only lures the model to learn from random noises. To overcome such problems, this paper categorizes reaction yields, which range from zero to one, into three types: high-yield, medium-yield, and low-yield. We refer to these labels as coarse reaction yield labels, or coarse labels in short.

We now look at how we adapt the SCL network to encode our coarse-label information into the fingerprint. SCL is a learning framework that extends the ideas of self-supervised contrastive learning to the supervised learning setting where label information is available. It was introduced to improve the quality of representations learned by deep learning models, with a particular focus on improving the robustness and generalization of the learned features. The loss function of our SCL network is
\begin{equation}
\mathcal{L}_{\text{SCL}} = \sum_{i=1}^{N} \frac{-1}{N_{y_i}} \sum_{\substack{j=1 \\ j \neq i \\ \xi (y_j) = \xi (y_i)}}^{N} \log \frac{\exp(\textbf{z}_i \cdot \textbf{z}_j / \tau)}{\sum_{\substack{k=1 \\ k \neq i}}^{N} \exp(\textbf{z}_i \cdot \textbf{z}_k / \tau)}
\end{equation}\label{SCL_loss}
where $\mathcal{L}_{SCL}$ denotes the supervised contrastive loss; $N$ is the number of reaction samples; $y_i$ is the reaction yield of the $i^{th}$ sample; $\xi (y_i)$ is the coarse label of $i^{th}$ sample; $N_{y_i}$ is the number of reaction samples that have the same coarse label as the sample $i$; $\mathbf{z}_i$ and $\mathbf{z}_j$ are the embeddings of the $i^{th}$ and $j^{th}$ samples, respectively; $\tau$ is a temperature scaling parameter that helps to control the separation of the classes.

An intuitive explanation of using the SCL network for our task is as follows. The loss function encourages the model to learn to minimize the distance between embeddings of examples with the same label and maximize the distance between embeddings of examples with different labels. In this way, SCL networks help to learn an embedding space where the representations of reactions are organized based on their coarse labels. This embedding space groups together reactions that have the same coarse label, while also ensuring that reactions that have different coarse labels are well-separated. This separation in the feature space means that the model learns to 1) bring closer the embeddings of samples with the same coarse label and 2) push apart embeddings of samples with different coarse labels. Such a feature space is very helpful in distinguishing the unique features of high-yield, medium-yield, and low-yield reactions.

One important thing we found is that the embedding constructed with the SCL scheme is a very efficient input for a variety of ML models. For example, in image processing research, \cite{Ref022} builds a convolutional neural network after extracting the embedding vector from the SCL network. In this regard, the post-SCL embedding vector is, in essence, a special reaction encoding that can be further processed with different ML models. Therefore, we follow the setting in \cite{bib24, bib25} and define the post-SCL embedding as a reaction fingerprint. 

In this paper, we tested three different ML models: random forest \cite{bib22}, tabular transformer \cite{Ref024}, and Xgboost \cite{bib23}. Implementations of the three ML models are packaged within Phase Two's API function. When calling the API, Chemist-X can select one of the three models for the execution of Phase Two via a user-friendly API interface.

\subsection{Phase Three: Automated Laboratory Execution with LLM-Supervised Interface Control}\label{subsection3D}
Many laboratory automation platforms lack user-accessible APIs for machine control, often relying on graphical user interfaces (GUIs) that require manual interactions such as mouse clicks and keyboard inputs \cite{bib034a,bib034b,bib034c}. To address this challenge, Chemist-X employs a novel LLM-supervised interface control framework that enables the automated execution of chemical experiments by replicating human interactions with laboratory software GUIs.\footnote{For a detailed Python implementation of our equipment control framework, we refer readers to our GitHub repository \cite{OurGithub}.}

To facilitate automated execution, Chemist-X is provided with a demonstration video that records a human operator interacting with the experimental software. This video serves as a reference for the LLM-based agent to understand and replicate the procedural workflow. The recorded demonstration typically consists of three key stages, including \textbf{1) Initialization}: Predefined actions such as launching the experiment control software and ensuring all necessary modules are loaded; \textbf{2) Reaction Condition Configuration}: Selection of appropriate reactants, solvents, catalysts, and substrates, along with specification of experimental parameters (e.g., temperature, reaction duration); and \textbf{3) Execution}: Verification of the experiment setup and initiation of the reaction.

\begin{figure}[htbp]
\centering
\includegraphics[width=0.98\textwidth]{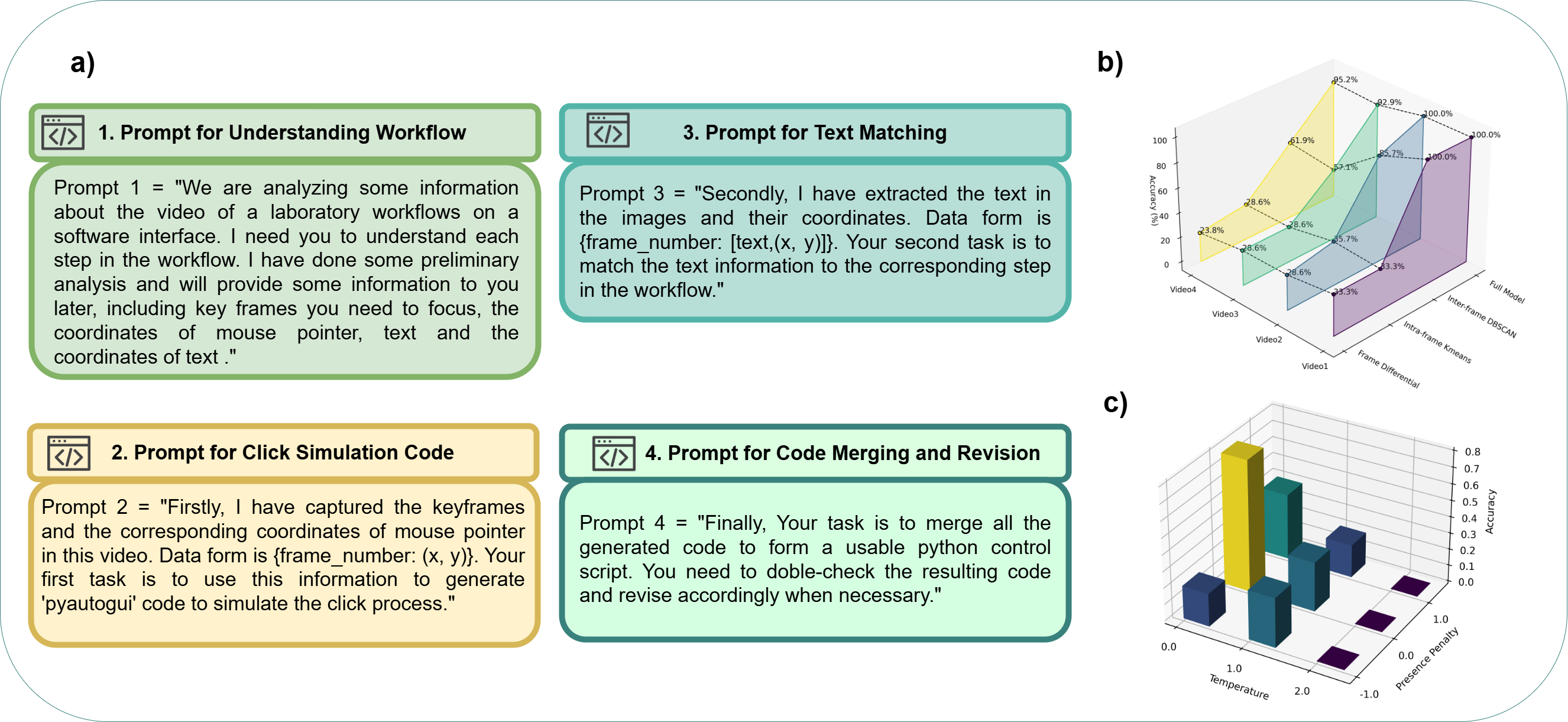}
\caption{Technical details about the LLM-supervised control script generation for wet-lab equipments. Figure a) presents our prompting framework. Figure b) gives the ablation study result for the mouse clicking algorithm. Figure c) shows the parameter sensitivity of the LLM-supervised system.}\label{Fig_PhaseThree}
\end{figure}

Among these stages, the initialization stage and the execution stage follow a fixed set of operations and can be automated by tracking and replicating mouse movements and click behaviors. In contrast, reaction condition configuration presents a greater challenge due to its dependency on specific reaction requirements. Figure \ref{Fig_PhaseThree}a provides a detailed illustration of how our novel multi-step prompt framework enables the LLM to understand, analyze, and replicate human interactions.

To begin with, we provide the LLM with the demonstration video alongside a general task description to ensure the model's comprehensive understanding of the workflow. A key frame detection algorithm is then applied in Step Two to filter out irrelevant frames (e.g., idle periods of the recorded video, during which time no human operation is conducted). For the detected key frames, we then apply a clustering algorithm to identify the mouse pointer locations to capture the user's clicking operations. The extracted information about the mouse's location is then passed to the LLM, which generates simulated click sequences (based on the \textit{pyautogui} Python package) for the fixed operation in Initialization and Execution stages. Note that engineering details of the key frame detecting algorithm and the mouse identification algorithm are not presented here for better scientific focus. Interested readers can refer to our open-source implementation in the GitHub repository \cite{OurGithub}.

Step Three handles the more complex Reaction Condition Configuration stage. Unlike the Initialization and Execution stages, which involve fixed operations, this stage requires adaptive decision-making based on the specific experiment being executed. The demonstration video only provides an example of human interaction, but the exact chemical parameters and selection choices may vary across different experiments. Therefore, directly mimicking recorded mouse clicks is insufficient. To address this, Chemist-X uses an optical character recognition (OCR) tool to extract textual information (e.g., reagent names, concentrations, dropdown menu items) from the software interface and then interprets the extracted text and matches this textual information with the desired reaction conditions (see the third prompt in Fig.\ref{Fig_PhaseThree}a for details). Following the matching results, the agent generates the associated control script (also based on the \textit{pyautogui} Python package) that can execute mouse click sequences to configure the reaction setup as required.  

Finally, Chemist-X combines the code generated in Step Two (i.e., for the Initialization and Execution stages) and Step Three (for the Reaction Condition Configuration stage) to obtain the desired control script and double-checks the code to fix potential errors.

\section{Conclusion}
This paper introduces Chemist-X, an innovative RAG-AI agent that harnesses the latest research breakthroughs in RAG technology and LLMs to transform the landscape of reaction condition recommendations in chemical synthesis. In general, this paper makes three significant contributions. First, with the ``search-analyze-recommend” pattern of expert chemists and the novel reaction fingerprint that integrates yield data, Chemist-X demonstrates remarkable proficiency in automating the RCO task. Second, the use of updated online databases as knowledge sources allows our agent to transcend the limitations of traditional AI systems, which are confined to the knowledge within their training data. Furthermore, the agent's ability to leverage cutting-edge CAD tools also enables the continuous evolution of the system. Third, the successful application of our RAG-AI agent to the Suzuki reaction example, validated through wet-lab experiments, underscores the practical impact of this research. With the above technical and experimental contributions, this research brings the vision of fully automated chemical discovery one step closer, marking a significant leap in the field of computer-aided chemistry.

\section{Extended Data}
\textit{Extended Data Fig. 1} presents the Code for PubChem API request, which is generated by Chemist-X. The generated code receives SMILES as input and leverages PubChem API to acquire similar molecules. 

\bibliography{sn-bibliography}
\end{document}